# Spin pinning and spin-wave dispersion in nanoscopic ferromagnetic waveguides

Q. Wang[1,*], B. Heinz[1,2,*], R. Verba[3], M. Kewenig[1], P. Pirro[1], M. Schneider[1], T. Meyer[1,4], B. Lägel[5], C. Dubs[6], T. Brächer[1], and A. V. Chumak[1,**]

[1]*Fachbereich Physik and Landesforschungszentrum OPTIMAS, Technische Universität Kaiserslautern, D-67663 Kaiserslautern, Germany*

[2]*Graduate School Materials Science in Mainz, Staudingerweg 9, 55128 Mainz, Germany*

[3]*Institute of Magnetism, Kyiv 03680, Ukraine*

[4]*THATec Innovation GmbH, Augustaanlage 23, 68165 Mannheim, Germany.*

[5]*Nano Structuring Center, Technische Universität Kaiserslautern, D-67663 Kaiserslautern, Germany*

[6]*INNOVENT e.V., Technologieentwicklung, Prüssingstraße 27B, 07745 Jena, Germany*

*: These authors have contributed equally to this work

**: Corresponding author: chumak@physik.uni-kl.de

**Abstract**

Spin waves are investigated in Yttrium Iron Garnet (YIG) waveguides with a thickness of 39 nm and widths ranging down to 50 nm, i.e., with aspect ratios thickness over width approaching unity, using Brillouin Light Scattering spectroscopy. The experimental results are verified by a semi-analytical theory and micromagnetic simulations. A critical width is found, below which the exchange interaction suppresses the dipolar pinning phenomenon. This changes the quantization criterion for the spin-wave eigenmodes and results in a pronounced modification of the spin-wave characteristics. The presented semi-analytical theory allows for the calculation of spin-wave mode profiles and dispersion relations in nano-structures.

Spin waves and their quanta, magnons, typically feature frequencies in the GHz to THz range and wavelengths in the micrometer to nanometer range. They are envisioned for the design of faster and smaller next generational information processing devices where information is carried by magnons instead of electrons [1-9]. In the past, spin-wave modes in thin films or rather planar waveguides with thickness-to-width aspect ratios $a_r = h/w \ll 1$ have been studied. Such thin waveguides demonstrate the effect of "dipolar pinning" at the lateral edges, and for its theoretical description the thin strip approximation was developed, in which only pinning of the much-larger-in-amplitude dynamic in-plane magnetization component is taken into account [10-15]. The recent progress in fabrication technology leads to the development of nanoscopic magnetic devices in which the width $w$ and the thickness $h$ become comparable [16-23]. The description of such waveguides is beyond the thin strip model of effective pinning, because the scale of nonuniformity of the dynamic dipolar fields, which is described as "effective dipolar boundary conditions", becomes comparable to the waveguide width. Additionally, both, in-plane and out-of-plane dynamic magnetization components, become involved in the effective dipolar pinning, as they become of comparable amplitude. Thus, a more general model should be developed and verified experimentally. In addition, such nanoscopic feature sizes imply that the spin-wave modes bear a strong exchange character, since the widths of the structures are now comparable to the exchange length [24]. A proper description of the spin-wave eigenmodes in nanoscopic strips which considers the influence of the exchange interaction, as well as the shape of the structure, is fundamental for the field of magnonics.

In this Letter, we discuss the evolution of the frequencies and profiles of the spin-wave modes in nanoscopic waveguides where the aspect ratio $a_r$ evolves from the thin film case $a_r \rightarrow 0$ to a rectangular bar with $a_r \rightarrow 1$. Yttrium Iron Garnet (YIG) waveguides with a thickness of 39 nm and widths ranging down to 50 nm are fabricated and the quasi-ferromagnetic resonance (quasi-FMR) frequencies within them are measured using microfocused Brillouin Light Scattering (BLS) spectroscopy. The experimental results are verified by a semi-analytical theory and micromagnetic simulations. It is shown that a critical waveguide width exists, below which the profiles of the spin-wave modes are essentially uniform across the width of the waveguide. This is fundamentally different from the profiles in state-of-the-art waveguides of micrometer [16-19] or millimeter sizes [25,26], where the profiles are non-uniform and pinned at the waveguide edges due to the dipolar interaction. In nanoscopic waveguides, the exchange interaction suppresses this pinning due to its dominance over the dipolar interaction and, consequently, the exchange interaction defines the profiles of the spin-wave modes as well as the corresponding spin-wave dispersion characteristics.

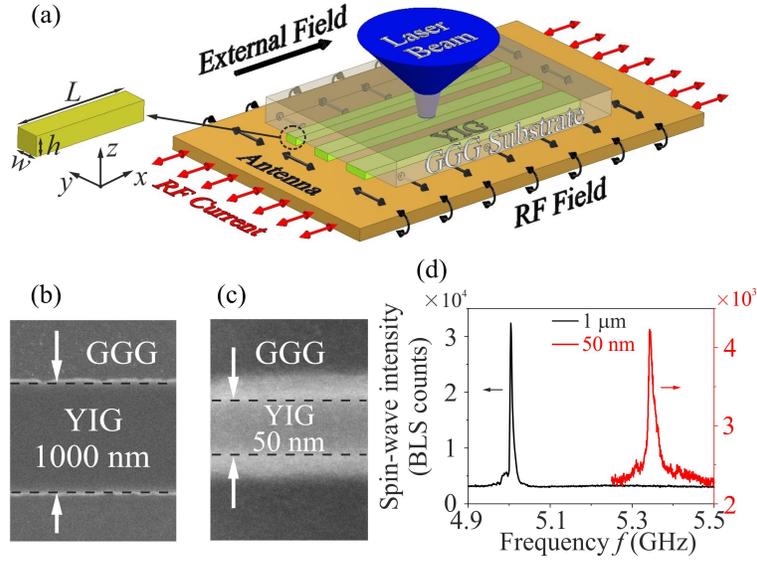

*Fig. 1 (a) Sketch of the sample and the experimental configuration: a set of YIG waveguides is placed on a microstrip line to excite the quasi-FMR in the waveguides. BLS spectroscopy is used to measure the local spin-wave dynamics. (b) and (c) SEM micrograph of a 1 µm and a 50 nm wide YIG waveguide of 39 nm thickness. (d) Frequency spectra for 1 µm and 50 nm wide waveguides measured for a respective microwave power of 6 dBm and 15 dBm.*

In the experiment and the theoretical studies, we consider rectangular magnetic waveguides as shown schematically in Fig. 1(a). In the experiment, a spin-wave mode is excited by a stripline that provides a homogeneous excitation field over the sample containing various waveguides etched from a $h = 39$ nm thick YIG film grown by liquid phase epitaxy [27] on Gadolinium Gallium Garnet (GGG). The widths of the waveguides range from $w = 50$ nm to $w = 1$ µm with a length of 60 µm. The waveguides are patterned by $Ar^+$ ion beam etching using an electron-beam lithographically defined Cr/Ti hard mask and are well separated on the sample in order to avoid dipolar coupling between them [9]. The waveguides are uniformly magnetized along their long axis by an external field $B$ ($x$-direction). Figure 1(b) and (c) show scanning electron microscopy (SEM) micrographs of the largest and the narrowest waveguide studied in the experiment. The intensity of the magnetization precession is measured by microfocused BLS spectroscopy [28] (see Supplementary Material S3 [29]) as shown in Fig. 1(a). Black and red lines in Fig. 1d show the frequency spectra for a 1 µm and a 50 nm wide waveguide, respectively. No standing modes across the thickness were observed in our experiment, as their frequencies lie higher than 20 GHz due to the small thickness. The quasi-FMR frequency is 5.007 GHz for the 1 µm wide waveguide. This frequency is comparable to 5.029 GHz, the value predicted by the classical theoretical model using the thin strip approximation [12-14, 34]. In contrast, the quasi-FMR frequency is 5.35 GHz for a 50 nm wide waveguide which is much smaller than the value of 7.687 GHz predicted by the same model. The reason is that the thin strip approximation overestimates the effect of dipolar pinning in waveguides with aspect ratio $a_r$ close to one, for which the nonuniformity of the dynamic dipolar fields is not well-localized at the waveguide edges. Additionally, in such nanoscale waveguides, the dynamic magnetization components become of the same order of magnitude and both affect the effective mode pinning, in contrast to thin waveguides, in which the in-plane magnetization component is dominant.

In order to accurately describe the spin-wave characteristic in nanoscopic longitudinally magnetized waveguides, we provide a more general semi-analytical theory which goes beyond the thin strip approximation. Please note that the theory is not applicable in transversely magnetized waveguides due to their more involved internal field landscape [16]. The lateral spin-wave mode profile $\mathbf{m}_{k_x}(y)$ and frequency can be found from [35,36]

$$-i\omega_{k_x}\mathbf{m}_{k_x}(y) = \boldsymbol{\mu} \times \left(\hat{\boldsymbol{\Omega}}_{k_x} \cdot \mathbf{m}_{k_x}(y)\right), \tag{1}$$

with appropriate exchange boundary conditions, which take into account the surface anisotropy at the edges (see Supplementary Material S1 [29]). Here, $\boldsymbol{\mu}$ is the unit vector in the static magnetization direction and $\hat{\boldsymbol{\Omega}}_{k_x}$ is a tensorial Hamilton operator, which is given by

$$\hat{\boldsymbol{\Omega}}_{k_x} \cdot \mathbf{m}_{k_x}(y) = \left(\omega_H + \omega_M \lambda^2 \left(k_x^2 - \frac{d^2}{dy^2}\right)\right)\mathbf{m}_{k_x}(y) + \omega_M \int \hat{\mathbf{G}}_{k_x}(y-y') \cdot \mathbf{m}_{k_x}(y')dy'. \tag{2}$$

Here, $\omega_H = \gamma B$, $B$ is the static internal magnetic field that is considered to be equal to the external field due to the negligible demagnetization along the $x$-direction, $\omega_M = \gamma\mu_0 M_s$, $\gamma$ is the gyromagnetic ratio. $\hat{\mathbf{G}}_{k_x}(y)$ is the Green's function (see Supplementary Material S1 [29]).

A numerical solution of Eq. (1) gives both, the spin-wave profiles $\mathbf{m}_{kx}$ and frequency $\omega_{kx}$. In the following, we will regard the ouf-of-plane component $m_z(y)$ to show the mode profiles representatively. The profiles of the spin-wave modes can be well approximated by sine and cosine functions. In the past, it was demonstrated that in microscopic waveguides, that fundamental mode is well fitted by the function $m_z(y)=A_0\cos(\pi y/w_{\text{eff}})$ with the amplitude $A_0$ and the effective width $w_{\text{eff}}$ [12,13]. This mode, as well as the higher modes, are referred to as 'partially pinned'. Pinning hereby refers to the fact that the amplitude of the modes at the edges of the waveguides is reduced. In that case, the effective width $w_{\text{eff}}$ determines where the amplitude of the modes would vanish outside the waveguide [9,12,23]. With this effective width, the spin-wave dispersion relation can also be calculated by the analytical formula [9]:

$$\omega_0(k_x) = \sqrt{\left(\omega_H + \omega_M\left(\lambda^2 K^2 + F_{k_x}^{yy}\right)\right)\left(\omega_H + \omega_M\left(\lambda^2 K^2 + F_{k_x}^{zz}\right)\right)}, \tag{3}$$

where $K = \sqrt{k_x^2 + \kappa^2}$ and $\kappa = \pi/w_{\text{eff}}$. The tensor $\hat{F}_{k_x} = \frac{1}{2\pi}\int_{-\infty}^{\infty} \frac{|\sigma_k|^2}{\tilde{w}} \hat{\mathbf{N}}_k dk_y$ accounts for the dynamic demagnetization, $\sigma_k = \int_{-w/2}^{w/2} m(y) e^{-ik_y y} dy$ is the Fourier-transform of the spin-wave profile across the width of the waveguide, $\tilde{w} = \int_{-w/2}^{w/2} m(y)^2 dy$ is the normalization of the mode profile $m_z(y)$.

In the following, the experiment is compared to the theory and to micromagnetic simulations. The simulations are performed using MuMax$^3$ [37]. The structure is schematically shown in Fig. 1(a). The following parameters were used: the saturation magnetization $M_s = 1.37 \times 10^5$ A/m and the Gilbert damping $\alpha = 6.41 \times 10^{-4}$ were extracted from the plain film via ferromagnetic resonance spectroscopy before patterning [38]. Moreover, a gyromagnetic ratio $\gamma = 175.86$ rad/(ns·T) and an exchange constant $A = 3.5$ pJ/m for a standard YIG film were assumed. An external field $B = 108.9$ mT is applied along the waveguide long axis (see Supplementary Material S2 [29]).

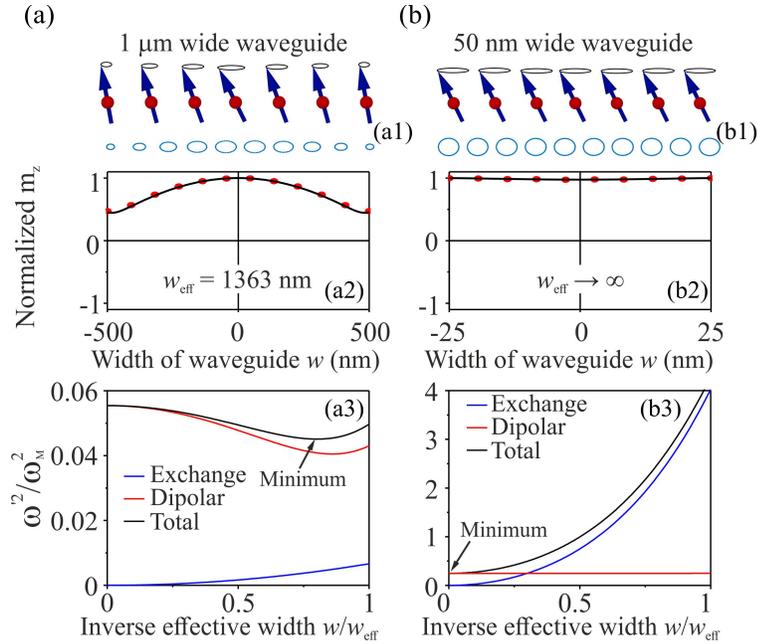

Fig. 2 Schematic of the precessing spins and simulated precession trajectories (ellipses in the second panel) and spin-wave profile $m_z(y)$ of the quasi-FMR. The profiles have been obtained by micromagnetic simulations (red dots) and by the quasi-analytical approach (black lines) for an (a) 1 µm and a (b) 50 nm wide waveguide. Bottom panel: Normalized square of the spin-wave eigenfrequency $\omega'^2/\omega_M^2$ as a function of $w/w_{\text{eff}}$ and the relative Dipolar and Exchange contributions.

The central panel of Fig. 2 shows the spin-wave mode profile of the fundamental mode for $k_x = 0$, which corresponds to the quasi-FMR, in a 1 µm (a2) and 50 nm (b2) wide waveguide which have been obtained by micromagnetic simulations (red dots) and by solving Eq. (1) numerically (black lines) (higher width modes are discussed in Supplementary Material S6 [29]). The top panels (a1) and (b1) illustrate the mode profile and the local precession amplitude in the waveguide. As it can be seen, the two waveguides feature quite different profiles of their fundamental modes: in the 1 µm wide waveguide, the spins are partially pinned and the amplitude of $m_z$ at the edges of the waveguide is reduced. This still resembles the cosine-like profile of the lowest width mode $n = 0$ that has been well established in investigations of spin-wave dynamics in waveguides on the micron scale [19,23,39] and that can be well-described by the simple introduction of a finite effective width $w_{\text{eff}} > w$ ($w_{\text{eff}} = w$ for the case of full pinning). In contrast, the spins at the edges of the narrow waveguide are completely unpinned and the amplitude of the dynamic magnetization $m_z$ of the lowest mode $n = 0$ is almost constant across the width of the waveguide, resulting in $w_{\text{eff}} \to \infty$.

To understand the nature of this depinning, it is instructive to consider the spin-wave energy as a function of the geometric width of the waveguide normalized by the effective width $w/w_{\text{eff}}$. This ratio corresponds to some kind of pinning parameter taking values in between 1 for the fully pinned case and 0 for the fully unpinned case. The system will choose the mode profile which minimizes the total energy. This is equivalent to a variational minimization of the spin-wave eigenfrequencies as a function of $w/w_{\text{eff}}$. To illustrate this, the lower panels of Figs. 2(a3) and (b3) show the normalized square of the spin-wave eigenfrequencies $\omega'^2 / \omega_M^2$ for the two different widths as a function of $w/w_{\text{eff}}$. Here, $\omega'^2$ refers to a frequency square, not taking into account the Zeeman contribution ($\omega_H^2 + \omega_H \omega_M$), which only leads to an offset in frequency. The minimum of $\omega'^2$ is equivalent to the solution with the lowest energy corresponding to the effective width $w_{\text{eff}}$. In addition to the total $\omega'^2$ (black), also the individual contributions from the dipolar term (red) and the exchange term (blue) are shown, which can only be separated conveniently from each other if the square of Eq. 3 is considered for $k_x = 0$. The dipolar contribution is non-monotonous and features a minimum at a finite effective width $w_{\text{eff}}$, which can clearly be observed for $w = 1$ µm. The appearance of this minimum, which leads to the effect known as "effective dipolar pinning" [13,14], is a results of the interplay of two tendencies: (i) an increase of the volume contribution with increasing $w/w_{\text{eff}}$, as for common Damon-Eshbach spin waves, and (ii) a decrease of the edge contribution when the spin-wave amplitude at the edges vanishes ($w/w_{\text{eff}} \to 1$). This minimum is also present in the case of a 50 nm wide waveguide (red line), even though this is hardly perceivable in Fig. 2(b3) due to the scale. In contrast, the exchange leads to a monotonous increase of frequency as a function of $w/w_{\text{eff}}$, which is minimal for the unpinned case, i.e., $w/w_{\text{eff}} = 0$ implying $w_{\text{eff}} \to \infty$, when all spins are parallel. In the case of the 50 nm waveguide, the smaller width and the corresponding much larger quantized wavenumber in the case of pinned spins would lead to a much larger exchange contribution than this is the case for the 1 µm wide waveguide (please note the vertical scales).

Consequently, the system avoids pinning and the solution with lowest energy is situated at $w/w_{eff} = 0$. In contrast, in the 1 µm wide waveguide, the interplay of dipolar and exchange energy implies that energy is minimized at a finite $w/w_{eff}$. The top panel of Fig. 2 (b1) shows an additional feature of the narrow waveguide: as the aspect ratio of the waveguides approaches unity, the ellipticity of precession, a well-known feature of micron-sized waveguides which still resemble a thin film [23,40], vanishes and the precession becomes nearly circular. Also, in nanoscale waveguides, the ellipticity is constant across the width, while in 1 µm wide waveguide it can be different at the waveguide center and near its edges. Please note that the pinning phenomena and ellipticity of precession also influence the spin-wave lifetime which is described in the Supplementary Material S5 [29].

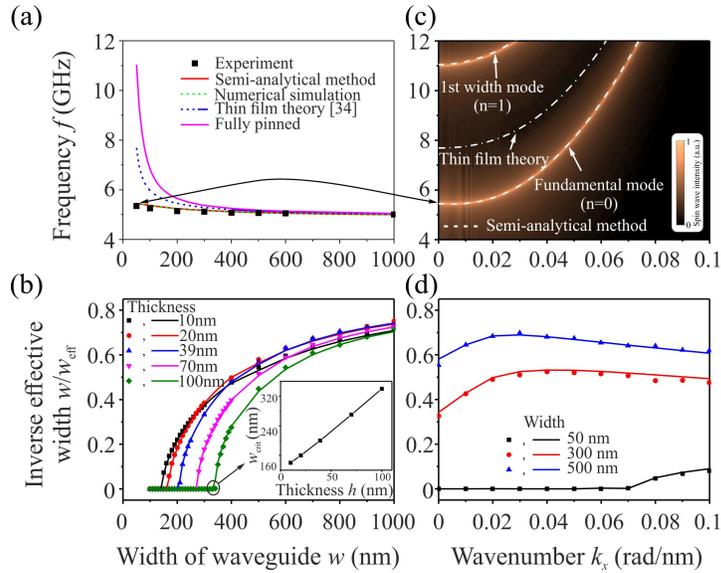

*Fig 3: (a) Experimentally determined resonance frequencies (black squares) together with theoretical predictions and micromagnetic simulations. (b) Inverse effective width $w/w_{eff}$ as a function of the waveguide width. The inset shows the critical width ($w_{crit}$) as a function of thickness h. (c) Spin-wave dispersion relation of the first two width modes from micromagnetic simulations (color-code) and theory (dashed lines). (d) Inverse effective width $w/w_{eff}$ as a function of the spin-wave wavenumber $k_x$ for different thicknesses and waveguide widths, respectively.*

As it is evident from the lower panel of Fig. 2, the pinning and the corresponding effective width have a large influence on the spin-wave frequency. This allows for an experimental verification of the presented theory, since the frequency of partially pinned spin-wave modes would be significantly higher than in the unpinned case. Black squares in Figure 3(a) summarize the dependence of the frequency of the quasi-FMR measured for different widths of the YIG waveguides. The magenta line shows the expected frequencies assuming pinned spins, the blue (dashed) line gives the resonance frequencies extrapolating the formula conventionally used for micron-sized waveguides [34] to the nanoscopic scenario, and the red line gives the result of the theory presented here, together with simulation results (green dashed line). As it can be seen, the experimentally observed frequencies can be well reproduced if the real pinning conditions are taken into account.

As has been discussed alongside with Fig. 2, the unpinning occurs when the exchange interaction contribution becomes so large that it compensates the minimum in the dipolar contribution to the spin-wave energy. Since the energy contributions and the demagnetization tensor change with the thickness of the investigated waveguide, the critical width below which the spins become unpinned is different for different waveguide thicknesses. This is shown in Fig. 3(b), where the inverse effective width $w/w_{eff}$ is shown for different waveguide thicknesses. Symbols are the results of micromagnetic simulations, lines are calculated semi-analytically. As can be seen from the figure, the critical width linearly increases with increasing thickness. This is summarized in the inset, which shows the critical width (i.e. the maximum width for which $w/w_{eff} = 0$) as a function of thickness. The critical widths for YIG, Permalloy, CoFeB and Heusler ($Co_2Mn_{0.6}Fe_{0.4}Si$) compound with different thicknesses are given in the Supplementary Material S9 [29]. A simple empirical linear formula is found by fitting the critical widths for different materials in a wide range of thicknesses:

$$w_{crit} = 2.2h + 6.7\lambda, \tag{4}$$

where $h$ is the thickness of waveguide and $\lambda$ is the exchange length. Please note that additional simulations with rough edges and a more realistic trapezoidal cross section of the waveguides are also provided in the Supplementary Material S7, S8 [29]. The results show that these effects have a small impact on the critical width.

Up to now, the discussion was limited to the special case of $k_x = 0$. In the following, the influence of finite wave vector will be addressed. The spin-wave dispersion relation of the fundamental ($n = 0$) mode obtained from micromagnetic simulations (color-code) together with the semi-analytical solution (white dashed line) are shown in Fig. 3(c) for the YIG waveguide of $w = 50$ nm width. The figure also shows the low-wavenumber part of the dispersion of the first width mode ($n = 1$), which is pushed up significantly in frequency due to its large exchange contribution. Both modes are described accurately by the quasi-analytical theory. As it is described above, the spins are fully unpinned in this particular case. In order to demonstrate the influence of the pinning conditions on the spin-wave dispersion, a hypothetic dispersion relation for the case of partial pinning is shown in the figure with a dash-dotted white line (the case of $w/w_{eff} = 0.63$ is considered that would result from the usage of the thin strip approximation [12]). One can clearly see that the spin-wave frequencies in this case are considerably higher. Figure 3(d) shows the inverse effective width $w/w_{eff}$ as a function of the wavenumber $k_x$ for three exemplary waveguide widths of $w = 50$ nm, 300 nm and 500 nm. As it can be seen, the effective width and, consequently, the ratio $w/w_{eff}$ shows only a weak nonmonotonic dependence on the spin-wave wavenumber in the propagation direction. This dependence is a result of an increase of the inhomogeneity of the dipolar fields near the edges for larger $k_x$, which increases pinning [14], and of the simultaneous decrease of the overall strength of dynamic dipolar fields for shorter spin waves. Please note that the mode profiles are not only important for the spin-wave dispersion. The unpinned mode profiles will also greatly improve the coupling efficiency between two adjacent waveguides [9, 41-43].

In conclusion, the quasi-FMR of individual wires with widths ranging from 1 μm down to 50 nm are studied by means of BLS spectroscopy. A difference in the quasi-FMR frequency between experiment and the prediction by the classical thin strip theory is found for 50 nm wide waveguides. A semi-analytical theory accounting for the nonuniformity of both in-plane and out-of-plane dynamic demagnetization fields is presented and is employed together with micromagnetic simulations to investigate the spin-wave eigenmodes in nanoscopic waveguides with aspect ratio $a_r$ approaching unity. It is found that the exchange interaction is getting dominant with respect to the dipolar interaction, which is responsible for the phenomenon of dipolar pinning. This mediates an unpinning of the spin-wave modes if the width of the waveguide becomes smaller than a certain critical value. This exchange unpinning results in a quasi-uniform spin-wave mode profile in nanoscopic waveguides in contrast to the cosine-like profiles in waveguides of micrometer widths and in a decrease of the total energy and, thus, frequency, in comparison to the fully or the partially pinned case. Our theory allows to calculate the mode profiles as well as the spin-wave dispersion, and to identify a critical width below which fully unpinned spins need to be considered. The presented results provide valuable guidelines for applications in nano-magnonics where spin waves propagate in nanoscopic waveguides with aspect ratios close to one and lateral sizes comparable to the sizes of modern CMOS technology.

**Acknowledgements:**
The authors thank Burkard Hillebrands and Andrei Slavin for valuable discussions. This research has been supported by ERC Starting Grant 678309 MagnonCircuits and by the DFG through the Collaborative Research Center SFB/TRR-173 "Spin+X" (Projects B01) and through the Project DU 1427/2-1. R. V. acknowledges support from the Ministry of Education and Science of Ukraine, Project 0118U004007.

# Supplementary Material

# Spin pinning and spin-wave dispersion in nanoscopic ferromagnetic waveguides

Q. Wang, B. Heinz, R. Verba, M. Kewenig, P. Pirro, M. Schneider, T. Meyer, B. Lägel, C. Dubs, T. Brächer, and A. V. Chumak

In the supplemental material, we first discuss the details of numerical solution of the eigenproblem in Section S1. The details of micromagnetic simulations and BLS measurements are discussed in Section S2 and S3. The dependence of the dynamic demagnetization tensor $F_{k_x}$ on the width of waveguides and the lifetime of spin waves in Section S4 and S5. The mode profiles of higher width modes are discussed in Section S6. In Section S7 and S8, we show the influence of a more realistic, trapezoidal cross-section for waveguides and of edge roughness on the spin pinning condition. In Section S9, we provide additional simulations with different materials and study the dependence of the critical width on the exchange length.

**S1. Numerical solution of the eigenproblem**

The eigenproblem Eq. (1) should be solved with proper boundary conditions at the lateral edges of the waveguide. Since we use a complete description of the dipolar interaction via Green's functions:

$$\hat{\mathbf{G}}_{k_x}(y) = \frac{1}{2\pi} \int_{-\infty}^{\infty} \hat{\mathbf{N}}_k e^{ik_y y} dk_y. \tag{S1}$$

Here,

$$\hat{\mathbf{N}}_k = \begin{pmatrix} \frac{k_x^2}{k^2} f(kh) & \frac{k_x k_y}{k^2} f(kh) & 0 \\ \frac{k_x k_y}{k^2} f(kh) & \frac{k_y^2}{k^2} f(kh) & 0 \\ 0 & 0 & 1 - f(kh) \end{pmatrix}, \tag{S2}$$

where $f(kh) = 1 - (1 - \exp(-kh))/(kh)$, $k = \sqrt{k_x^2 + k_y^2}$ and it is assumed that the waveguides are infinitely long.

The boundary conditions account for exchange interaction and surface anisotropy (if any) only, and read [1]:

$$\mathbf{m} \times \left( \mu_0 M_s \lambda^2 \frac{\partial \mathbf{m}}{\partial \mathbf{n}} - \nabla_\mathbf{M} E_a \right) = 0, \tag{S3}$$

where **n** is the unit vector defining an inward normal direction to the waveguide edge, and $E_a(\mathbf{m})$ is the energy density of the surface anisotropy. In the studied case of a waveguide magnetized along its long axis, the conditions (S1) for dynamic magnetization components can be simplified to:

$$\pm \frac{\partial m_y}{\partial y} + dm_y \bigg|_{y=\pm w/2} = 0, \quad \frac{\partial m_z}{\partial y}\bigg|_{y=\pm w/2} = 0, \tag{S4}$$

where $d = -2K_s / (\mu_0 M_s^2 l^2)$ is the pinning parameter [2] and $K_s$ is the constant of surface anisotropy at the waveguide lateral edges. More complex cases like, e.g., diffusive interfaces can be considered in the same manner [3].

For the numerical solution of Eq. (1) it is convenient to use finite element methods and to discretize the waveguide into $n$ elements of the width $\Delta w = w/n$, where $w$ is the width of waveguide. The discretization step should be at least several times smaller than the waveguide thickness and the spin-wave wavelength $2\pi/k_x$ for a proper description of the magneto-dipolar fields. The discretization transforms Eq. (1) into a system of linear equations for magnetizations $\mathbf{m}_j$, $j = 1,2,3,\ldots n$:

$$i\boldsymbol{\mu} \times \left( \left( \omega_H + \omega_M \lambda^2 k_x^2 \right) \mathbf{m}_j - \omega_M \lambda^2 \frac{\mathbf{m}_{j-1} - 2\mathbf{m}_j + \mathbf{m}_{j+1}}{\Delta w^2} + \omega_M \sum_{j'=1}^{n} \hat{\mathbf{G}}_{j-j'} \cdot \mathbf{m}_{j'} \right) = \omega \mathbf{m}_j, \tag{S5}$$

where dipolar interaction between the discretized elements is described by

$$\hat{\mathbf{G}}_{k_x,j}(y) = \frac{1}{\Delta w} \int_{-\Delta w/2}^{\Delta w/2} dy \int_{-\Delta w/2}^{\Delta w/2} dy' \hat{\mathbf{G}}_{k_x}(y - y' - j\Delta w). \tag{S6}$$

The direct use of Eq. (S6) is complicated since the Green's $\hat{\mathbf{G}}_{k_x}(y)$ function is an integral itself. Using Fourier transform (FFT) it can be derived as

$$\hat{\mathbf{G}}_{k_x,j}(y) = \frac{\Delta w}{2\pi} \int \mathrm{sinc}(k_y \Delta w/2) \hat{\mathbf{N}}_k e^{ik_y j\Delta w} dk_y, \tag{S7}$$

which can be easily calculated, especially using FFT. Equation (S5) is, in fact, a $2n$-dimensional linear algebraic eigenproblem (since $\mathbf{m}_j$ is a 2-component vector), which is solved by standard methods. The values $\mathbf{m}_0$ and $\mathbf{m}_{n+1}$ in Eq. (S5) are determined from the boundary conditions (S4). In particular, for negligible anisotropy at the waveguide edges one should set $\mathbf{m}_0 = \mathbf{m}_1$ and $\mathbf{m}_{n+1} = \mathbf{m}_n$.

**S2. Micromagneitc simulations and data post-processiong**

The micromagnetic simulations were performed by the GPU-accelerated simulation program Mumax$^3$ to calculate the space- and time-dependent magnetization dynamics in investigated structures using a finite-difference discretization. The material parameters were given in the main text. There were three steps involved in the calculation of the spin-wave dispersion curve: (i) The external field was applied along the waveguide, and the magnetization was relaxed to a stationary state (ground state). (ii) A sinc field pulse $b_y = b_0 \mathrm{sinc}(2\pi f_c t)$, with oscillation field $b_0 = 1$ mT and cutoff frequency $f_c = 10$ GHz, was used to excite a wide range of spin waves. (iii) The spin-wave dispersion relations were obtained by performing the two-dimensional Fast Fourier Transformation of the time- and space-dependent data. Furthermore, the spin-wave width profiles were extracted from the $m_z$ component across the width of the waveguides using a single frequency excitation.

### S3. Microfocused Brillouin Light Scattering (BLS) spectroscopy measurements

BLS is a unique technique for measuring the spin-wave intensities in frequency, space, and time domains. It is based on inelastic light scattering of the incident laser beam from magnetic materials. In our measurements, a laser beam of 457 nm wavelength and a power of 1.8 mW is focused through the transparent GGG substrate on the center of the respective individual waveguide using a ×100 microscope objective with a large numerical aperture (NA=0.85). The effective spot-size is 350 nm. The scattered light was collected and guided into a six-pass Fabry-Pérot interferometer to analyze the frequency shift.

### S4. Width dependence of the dynamic demagnetization tensor

In the manuscript, we have demonstrated the change of the spin-wave pinning condition and the quasi-ferromagnetic resonance frequency in nanoscopic waveguides. Here, we investigate how the dynamic demagnetization tensor depends on the width of the waveguide. Neglecting the exchange term in Eq. (3) in the manuscript, the quasi-ferromagnetic resonance frequency can be expressed as:

$$\omega(k=0) = \sqrt{\omega_H^2 + \omega_H \omega_M \left( F_0^{yy} + F_0^{zz} \right) + \omega_M^2 F_0^{yy} F_0^{zz}}, \tag{S8}$$

where $F_0^{yy}$ and $F_0^{zz}$ are the $y$ and $z$ components of the demagnetization tensors. Equation (S8) clearly shows that the quasi-ferromagnetic resonance frequency will depend on the sample geometry as it determines the demagnetization tensor. The dependence of the demagnetization tensor components $F_0^{yy}$ and $F_0^{zz}$ on the width of the waveguide for a fixed thickness of 39 nm is shown in Fig. S1. The $y$ component of the demagnetization tensor $F_0^{yy}$ is close to zero in a wide range ($w > 2$ μm). However, $F_0^{yy}$ strongly increases with decreasing width of the waveguide and finally $F_0^{yy} = F_0^{zz} = 0.5$ for a 39 nm wide waveguide ($a_r=1$). The change of the demagnetization tensor indicates that the spin precession trajectory changes from elliptic to circular. Also, it means that in narrow waveguides with aspect ratio $a_r < 3$-5 dynamic magnetization components become of the same order and nonuniformity of both $yy$ and $zz$ dynamic demagnetization fields affects the effective pinning of the spin-wave modes. This was disregarded in the commonly used thin waveguide theory.

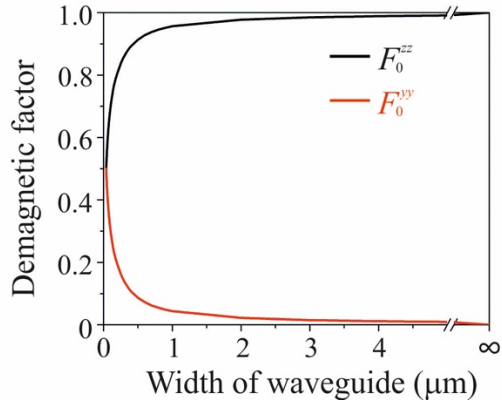

*Fig. S1 The dynamic demagnetization tensor components $F_0^{yy}$ and $F_0^{zz}$ as a function of the width of the waveguide for a fixed thickness of 39 nm.*

## S5. Spin-wave lifetime in magnetic nanostructures

The relaxation lifetime $\tau$ of uniform the procession mode in an infinite medium (without inhomogeneous linewidth $\Delta B_0$) is simply defined as $\tau = 1/(\alpha\omega)$, where $\omega$ is the angular frequency of the spin wave and $\alpha$ is the damping. However, the dynamic demagnetizing field has to be taken into account in finite spin-wave waveguide. The lifetime can be found by the phenomenological model [4-6]:

$$\tau = \left(\alpha\omega \frac{\partial \omega}{\partial \omega_H}\right)^{-1}. \tag{S9}$$

The dispersion relation has been shown in the manuscript (Eq. (3)). The demagnetization tensors are independent of $\omega_H$. Differentiating Eq. (3) yields the lifetime as

$$\tau = \left(\frac{1}{2}\alpha\left(2\omega_H + 2\omega_M \lambda^2 K^2 + \omega_M\left(F_{k_x}^{zz} + F_{k_x}^{yy}\right)\right)\right)^{-1}. \tag{S10}$$

This formula clearly shows that the lifetime of the uniform precession ($k_x$=0) depends only on the sum of the dynamic *yy* and *zz* components of demagnetization tensors.

Figure S2(a) shows the cross-section, spin precession trajectory (red line) and the dynamic components of the demagnetization tensors of different sample geometries. The spin precession trajectory changes from elliptic for the thin film ($a_r \ll 1$) to circular for the nanoscopic waveguide ($a_r$=1). The spin precession trajectory in the bulk material is also circular (in the geometry when spin waves propagate parallel to the static magnetic field, the same geometry as studied for nanoscale waveguides).

The dependence of the lifetime on the wavenumber is shown in Fig. S2(b) for YIG with a damping constant $\alpha = 2\times10^{-4}$. The inhomogeneous linewidth is not taken account. The lifetime of uniform precession ($k_x$=0) for the bulk material is much large than that in the thin film and nanoscopic waveguide, another consequence of the absence of dynamic demagnetization in the bulk ($F_0^{yy} = F_0^{zz} = 0$). Moreover, the lifetimes of the uniform precession ($k_x$=0) for a thin film (red line) and for a nanoscopic waveguide (black line) have the same value, because the lifetime depends only on the sum of the two components, which is the same for both cases.

Moreover, the *yy* and *zz* components of the demagnetization tensor decrease with an increase of the spin-wave wavenumber (instead, the *xx* component, which does not affect the spin wave dynamic in our geometry, increases). The lifetime is inversely proportional to the square of the wavenumber and the sum of the dynamic demagnetization components. In the exchange region, the lifetime is, thus, dominated by the wavenumber. Therefore, the lifetimes for short-wave spin-waves are nearly the same for the three different geometries.

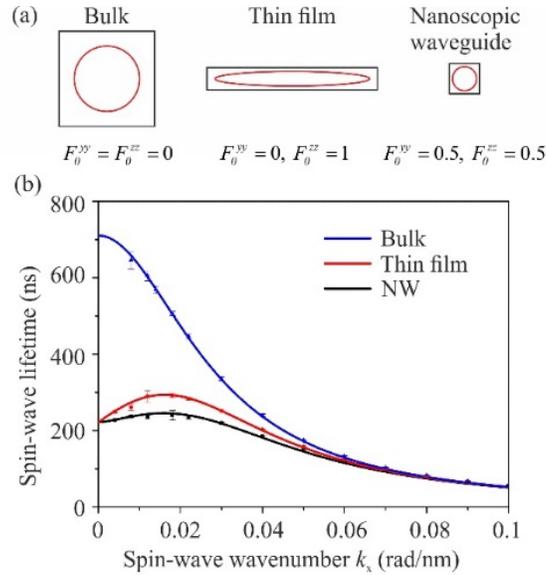

*Fig. S2 (a) The spin precession trajectories (red lines) and the components of demagnetization tensor $F_0^{yy}$ and $F_0^{zz}$ for different sample geometries. (b) The spin-wave lifetime as a function of spin-wave wavenumber. The lines and dots are obtained from Eq. (S10) and micromagnetic simulation, respectively.*

## S6. Profiles of higher width modes

In the manuscript, only the profile of the fundamental mode ($n = 0$) has been discussed. The mode profiles of higher width modes are shown in Fig. S3. It is clear to see that the spins are also fully unpinned at the edges for the higher width modes in a 50 nm wide waveguide. In contrast, the precession amplitude of the spins at the edges of a 1 μm waveguide increases with increasing mode number and is already almost equal to the maximum amplitude in the center of waveguide for the second width mode ($n = 2$). This change is a results of the increase of the exchange contribution for higher width modes.

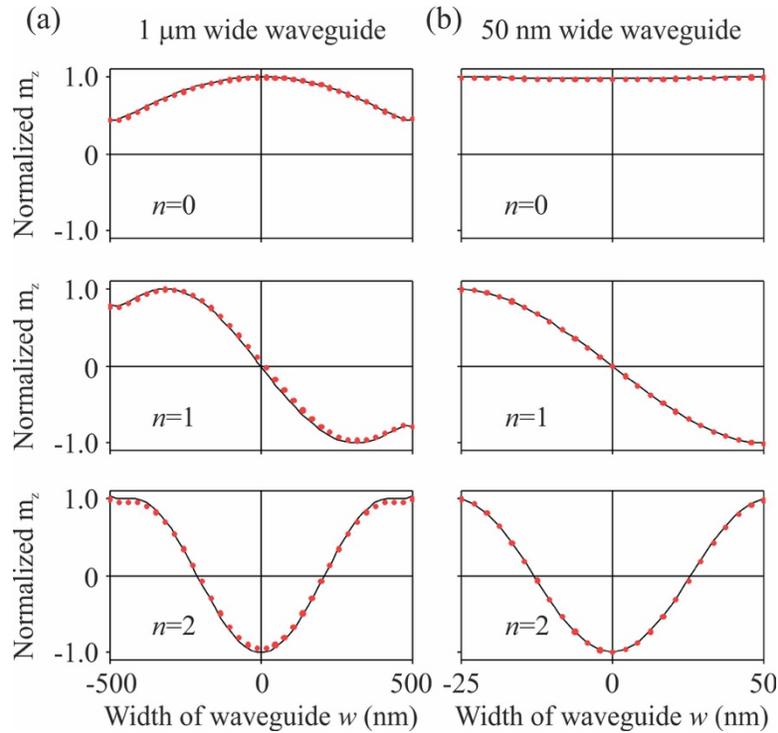

*Fig. S3 The spin-wave profile of the z component of the dynamic magnetization $m_z$ in the three lowest width modes obtained by micromagneitc simulation (black solid lines) and numerical calculation (red dots) for (a) 1 μm and (b) 50 nm wide waveguides.*

## S7. Influence of a trapezoidal form

A perfect rectangular form is not achievable in the experiment due to the involved patterning technique. As a result of the etching, the cross-section of the waveguides is always slightly trapezoidal. In this section, the influence of such a trapezoidal form on the spin pinning conditions is studied. In our experiment, the trapezoidal edges extent for around 20 nm on both sides for all the patterned waveguides. We performed additional simulation on waveguides with such trapezoidal edges. The simulated cross-section is shown in the top of Fig. S4. The thickness of the waveguide is divided into 5 layers with different widths ranging from 90 nm to 50 nm. The steps at the edges are hard to be avoided due to the finite difference method used in MuMax$^3$. The spin-wave profiles in the different $z$-layers are shown at the bottom of Fig. S4. The results clearly show that the spin-wave profiles are fully unpinned along the entire thickness. This is due to the fact that the largest width (90 nm) is still far below the critical width. Hence, the influence of the trapezoidal form of the waveguide on the spin pinning condition is negligible for very narrow waveguides. For large waveguides, it also does not have a large impact as the ratio of the edge to the waveguide area becomes close to zero. Quantitatively, the quasi-ferromagnetic resonance frequency in a 50 nm wide waveguide decreases from 5.45 GHz for the rectangular shape to 5.38 GHz for the trapezoidal form due to the increase of the averaged width which, in fact, even closer to the experiment results (5.35 GHz).

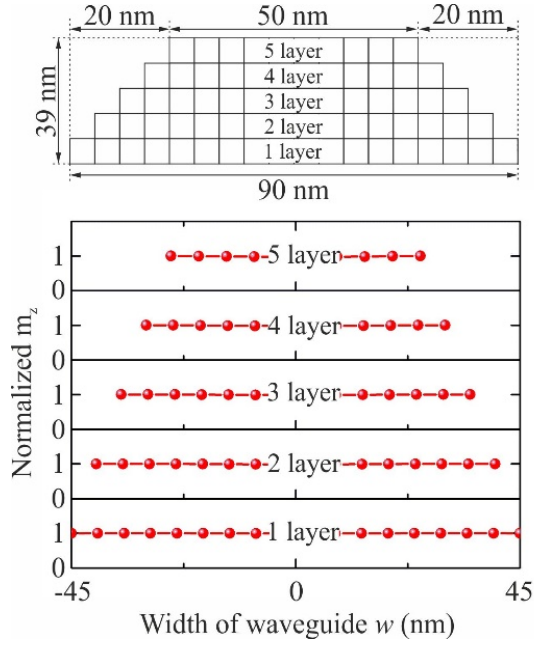

Fig. S4 Top: The cross section of trapezoid waveguide in the simulation. Bottom: The normalized spin-wave profile for different layers.

The inverse effective width $w/w_{\text{eff}}$ as a function of the width of the waveguides is simulated for a trapezoidal and a rectangular form and the result is shown in Fig. S5. Here, the width is defined by the minimal width for the trapezoidal form, i.e., the width of the top layer. In the case of trapezoidal form, the inverse effective width is averaged over all 5 layers. The critical width slightly decreases from 200 nm for the rectangular cross-section to 180 nm for the trapezoidal form due to the increase of the averaged width. The difference between the inverse effective widths decreases with increasing width of the waveguide and vanishes when the width is larger than 300 nm.

Furthermore, it should be noted that the results of the multilayer simulations demonstrate that the assumption of a uniform dynamic magnetization distribution across the thickness that is used in our analytical theory and micromagnetic simulations featuring only one cell in the z dimension is valid.

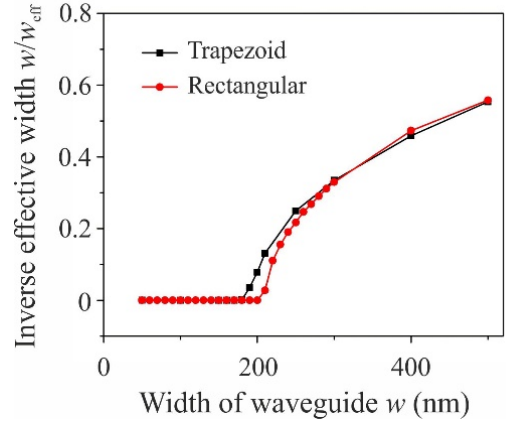

*Fig. S5 The inverse effective width $w/w_{eff}$ as a function of the width of waveguide for trapezoidal and rectangular form.*

## S8. Influence of edge roughness

Perfectly smooth edges are also hard to obtain in the experiment. We have also considered the influence of edge roughness on the spin pinning. We performed additional simulations on waveguides with rough boundaries for a fixed thickness of 39 nm. 5 nm (for 50 nm to 100 nm wide waveguides) or 10 nm (for 100 nm to 1000 nm wide waveguides) wide rectangular nonmagnetic regions with a random length are introduced randomly on both sides of the waveguides to act as defects. The introduction of roughness results in a slight increase of the critical width from 200 nm to 240 nm, as is shown in Fig. S6(a). These results demonstrate that edge roughness does not have a large influence on spin pinning condition.

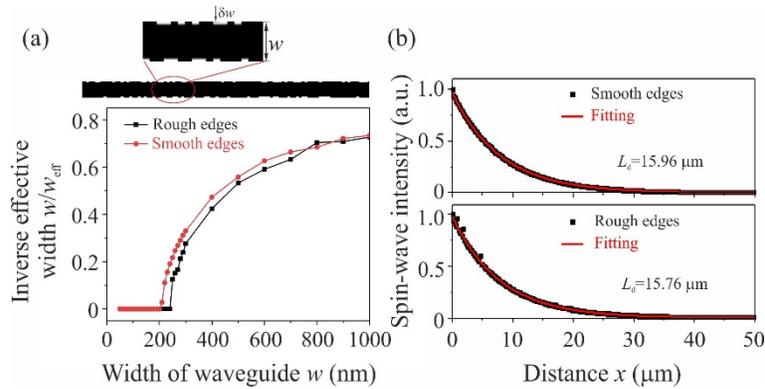

*Fig. S6 (a) Top: the schematic of rough waveguide and close-up image. Bottom: Inverse effective width $w/w_{eff}$ as a function of the waveguide width for rough and smooth edges. (b) The normalized spin-wave intensity as a function of propagation length for smooth and rough edges waveguide of 50 nm width.*

Additional simulations are performed to study the influence of a rough edge on the propagation length of spin waves with frequency 6.16 GHz ($k_x$=0.03 rad/nm). Figure S6(b) shows the normalized spin-wave intensity as a function of propagation length for smooth and rough edged waveguide of 50 nm width. The decay length slightly decreases from 15.96 μm for smooth edges to 15.76 μm for rough edges. Since the spins in nanoscopic waveguides are already unpinned, the effect of such an edge roughness is not too important anymore and the propagation length is essentially unaffected.

**S9. Critical width for different materials**

Figure S7 shows the inverse effective width $w/w_{eff}$ as a function of the waveguide width for typical materials used in magnonics. The inset shows the critical width ($w_{crit}$) as a function of exchange length $\lambda$ for different thicknesses. The critical width is proportional to the exchange length $\lambda$. A simple empirical linear formula is found by fitting the critical widths for different materials in a wide range of thicknesses to estimate the critical width: $w_{crit} = 2.2h + 6.7\lambda$, where $h$ is the thickness of the waveguide and $\lambda$ is the exchange length given by $\lambda = \sqrt{2A/(\mu_0 M_s^2)}$ with the exchange constant $A$, the vacuum permeability $\mu_0$, and the saturation magnetization $M_s$.

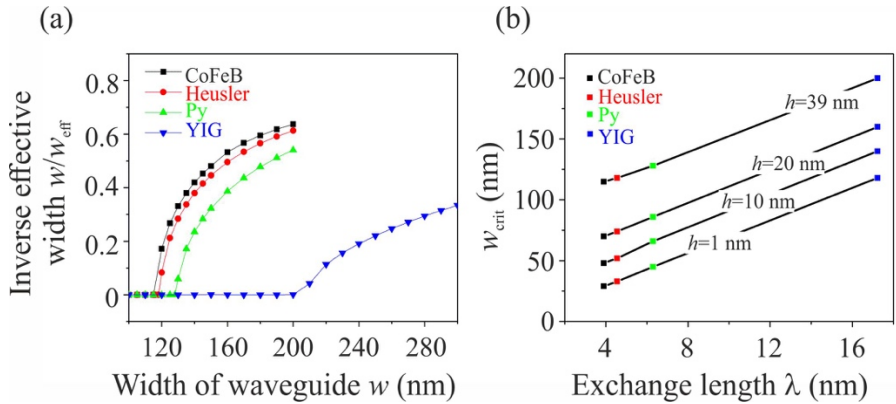

*Fig. S7 (a) Inverse effective width $w/w_{eff}$ as a function of waveguide width for different materials at fixed thickness of 39 nm. (b) the critical width ($w_{crit}$) as a function of exchange length $\lambda$ for different thicknesses.*